\documentclass[aps,prd,reprint,superscriptaddress, footinbib]{revtex4-1}

\usepackage{amsmath,amssymb, amsthm,amstext}
\usepackage{natbib}
\usepackage{graphicx}
\usepackage{color}
\usepackage{array, enumerate}
\usepackage{bm}
\usepackage{multirow}
\usepackage[breaklinks,colorlinks,citecolor=blue]{hyperref}
\usepackage{braket}
\usepackage{txfonts}

\def\be{\begin{equation}}
\def\ee{\end{equation}}
\def\apjl{ApJL}
\def\apj{ApJ}
\def\jcap{JCAP}

\begin{document}
\title{Improving the Detection Sensitivity to Primordial Stochastic Gravitational Waves with Reduced  Astrophysical Foregrounds}

\author{Zhen Pan}
\email{zpan@perimeterinstitute.ca}
\affiliation{Tsung-Dao Lee Institute, Shanghai Jiao-Tong University, Shanghai, 520 Shengrong Road, 201210, People’s Republic of China}
\affiliation{School of Physics \& Astronomy, Shanghai Jiao-Tong University, Shanghai, 800 Dongchuan Road, 200240, People’s Republic of China}
\affiliation{Perimeter Institute for Theoretical Physics, Ontario, N2L 2Y5, Canada}
\author{Huan Yang}
\email{hyang@perimeterinstitute.ca}
\affiliation{Perimeter Institute for Theoretical Physics, Ontario, N2L 2Y5, Canada}
\affiliation{University of Guelph, Guelph, Ontario N1G 2W1, Canada}

\date{\today}
\begin{abstract}
  One of the primary targets of third-generation (3G) ground-based gravitational wave (GW) detectors is detecting the stochastic GW background (SGWB)
  from early universe processes. The astrophysical foreground from compact binary mergers will be
  a major contamination to the background, which must be reduced to high precision to enable the detection of primordial background.
In this work, we revisit the limit of foreground reduction computed in previous studies, point out potential problems in previous foreground cleaning methods
and propose a novel cleaning method subtracting the approximate signal strain and removing the average residual power.
  With this method, the binary black hole foreground is reduced
  with fractional residual energy density below $10^{-4}$ for frequency $f\in (10, 10^2)$ Hz, below $10^{-3}$ for frequency $f\in (10^2, 10^3)$ Hz
  and below the detector sensitivity limit for all relevant frequencies in our simulations.  Similar precision is achieved to clean the foreground from
  binary neutron stars (BNSs) that are above the detection threshold, so that the residual foreground  is dominated by  sub-threshold BNSs,
  which will be the next critical  problem to solve for detecting the primordial SGWB in the 3G era.
\end{abstract}
\maketitle

\section{Introduction}
Primordial stochastic gravitational wave background (SGWB) from various physical processes from the early universe has been investigated,
including inflation \cite{Guth1981,Linde1982} and preheating \cite{Allahverdi2010,Amin2015},   first-order phase transitions \cite{Turner1990,Turner1992,Kamionkowski1994,Kosowsky1992,Caprini2008,Cutting2018}
and cosmic strings \cite{Vilenkin1981,Hogan1984,Caldwell1992,Hindmarsh1995,Vilenkin2000,Buchmuller2021} (see \cite{Allen1997b,Chiara2016,Cai2017,Caprini2018,Christensen2019,Renzini2022} for more complete reviews).
Measuring the primordial SGWB at different frequencies will open a unique window to the universe at the earliest moments.
Therefore the primordial SGWB detection has been one of the primary targets for gravitational wave (GW) detectors in different frequency bands, including pulsar timing arrays \cite{Hobbs2010,McLaughlin2013,Hobbs2013,Lentati2015}, spaceborne GW detectors \cite{Thorpe2019,Mei2021} and ground based detectors \cite{AdvLIGO2015,AdvVirgo2015,KAGRA2013}.

In addition to the primordial SGWB, GWs from various astrophysical sources are much better understood and measured \cite{1stGW,1stBNS,LVCO1O2,LVKCO3}.
In the sensitive band of ground based detectors, compact binaries, including binary black holes (BBHs), black hole-neutron star binaries  (BHNSs) and binary neutron stars
(BNSs) are the dominant source of astrophysical foreground \cite{Zhu2011,Zhu2013,Marassi2011,Wu2012}. In order to improve the  sensitivity of probing the primordial background, the effect of astrophysical foreground must be reduced. Cleaning the astrophysical foreground with the residual foreground energy density below the detector sensitivity limit has been argued to be  possible in the era of third-generation (3G) ground-based detectors, e.g., Einstein Telescope (ET) \cite{Maggiore2020} and  Cosmic Explorer (CE) \cite{Reitze2019,Evans2021,Srivastava2022}, which are so sensitive that almost all compact binary mergers are expected to be detected
and  subtracted out \cite{Regimbau2017}. However, recently Zhou et al. \cite{Zhou2022,Zhou2022b} pointed out the straightforward event subtraction proposed in \cite{Regimbau2017} only removes $\sim 50\%$ of the foreground noise. The resulting   astrophysical foreground is way above the detector sensitivity, which poses severe challenges of detecting primordial GWs \footnote{Recently, Zhong et al. \cite{Zhong2022} proposed a cleaning method by notching the individually resolved compact binary signals in the time-frequency domain instead, and they found the residual improves substantially but still well exceeds  the detector sensitivity limit.}.

In this work, we will show that the astrophysical foreground can be cleaned by orders of magnitude, with  a novel cleaning method detailed later.
Applying this method to 3G detectors, the BBH foreground and the foreground from individually resolved BNSs can be reduced to be well below the detector sensitivity limit.
As a result, the residual foreground is expected to be dominated by sub-threshold BNSs. Notice that the noise reduction of  sub-threshold events may be achieved following a  Bayesian  framework, as discussed in \cite{Drasco2003,Smith2019,Biscoveanu2020}. One subtlety may be that this method requires enormous computational cost, and it is more susceptible to non-Gaussian noise in detectors.
For space-borne detectors, it has also been shown that astrophysical foreground cleaning may benefit from multi-band observations \cite{Pan2020}.

In this paper, we use geometrical units $G=c=1$ and we assume a flat $\Lambda$CDM cosmology with $H_0=70$ km/s/Mpc, $\Omega_{ \Lambda}=0.7$ and $\Omega_{\rm m} = 0.3$.

\vspace{0.2 cm}
\section{Stochastic GW Foreground from Compact Binary Mergers}

The energy density of stochastic GWs per logarithmic frequency
is related to its power spectrum density (PSD) $H(f)$ by (our notation is different from that in \cite{Allen1997,Allen1999} by a factor $8\pi$)
\be\label{eq:omega}
\begin{aligned}
  \Omega_{\rm GW}(f)
  := \frac{1}{\rho_{\rm crit}} \frac{d\rho_{\rm GW}}{d \ln f} = \frac{4\pi^2}{3H_0^2}f^3 H(f)\ ,
\end{aligned}
\ee
with $\rho_{\rm crit}:=3 H_0^2/8\pi$ the critical energy density to close the universe.
 For the astrophysical foreground of compact binaries,
the PSD can be calculated as (see e.g., \cite{Allen1999,Phinney2001,Pan2020} for derivation)
\be\label{eq:HA}
 H(f) = \frac{1}{T}\sum_i  \left(|h_+(f)|^2+|h_\times(f)|^2 \right)_i   \ ,
\ee
where the index $i$ runs over all binaries in the universe that merger within the observation time span $(0, T)$
, and $h_{+,\times}$ are the two polarizations of incoming GWs. Driven by the incoming GWs,
the detector strain responds as
\be
  h(f) = F_+(\theta, \phi, \psi) h_+(f) + F_\times(\theta, \phi, \psi) h_\times(f)\ , \\
\ee
where the attena pattern $F_{+,\times}$ depend on the source sky location ($\theta,\phi$) and the source polarization angle $\psi$.
In terms of the detector strain, the PSD is expressed as
\be
  H(f) = \frac{2}{\braket{F_+^2}+\braket{F_\times^2}}\frac{1}{T}\sum_i |h(f)|^2_i = \frac{5}{T}\sum_i |h(f)|^2_i\ ,
\ee
where $\braket{}$ is an average over the three attena pattern dependent angles.
It is known that $\braket{F_+^2}=\braket{F_\times^2} = 1/5$
for LIGO/Virgo/KAGRA (LVK) like L-shape interferometers
and $\braket{F_+^2}=\braket{F_\times^2} = 3/20$
for LISA/ET like triangle-shape interferometers (see e.g., \cite{Sathyaprakash2009} for details).
We have chosen a L-shape interferometer as the reference detector in the second equal sign of the equation above.

A typical non-precessing BBH waveform depends on 7 model parameters
$h_{+,\times}(\mathcal M_z, M_z, \chi, t_c, \phi_c, \iota, D_{\rm L})$:
the (redshifted) chirp mass $\mathcal M_z =(1+z)\mathcal M$, the (redshifted) total mass $M_z=(1+z) M$,
the effective spin $\chi$, the coalescence time $t_c$ (arriving at a detector), the coalescence phase $\phi_c$, the inclination angle $\iota$, the luminosity distance $D_{\rm L}$.
In phenomenon waveform models \cite{Ajith2011}, the two polarizations are formulated as
\be
\begin{aligned}
  h_+(f)&=\frac{1+\cos^2\iota}{2}\sqrt{\frac{5}{24}}\frac{  \mathcal M_z^{5/6} f^{-7/6}}{\pi^{2/3} D_{\rm L}} A_{\rm phen}(f) e^{i[2\pi f t_c + \phi_c + \Psi_{\rm phen}(f)]}\ , \\
  h_\times(f)&=-i\cos\iota\sqrt{\frac{5}{24}}\frac{  \mathcal M_z^{5/6} f^{-7/6}}{\pi^{2/3} D_{\rm L}} A_{\rm phen}(f) e^{i[2\pi f t_c + \phi_c + \Psi_{\rm phen}(f)]}\ ,
\end{aligned}
\ee
where the waveform dependence on the intrinsic binary parameters is encoded in functions $A_{\rm phen}(f;\mathcal M_z, M_z, \chi)$ and $\Psi_{\rm phen}(f;\mathcal M_z, M_z, \chi)$.
In this work, we will use the simple PhenomB waveform model \cite{Ajith2011}
(the foreground cleaning results have little change if  other waveform models are applied instead, e.g., PhenomC or PhenomD \cite{Santamar2010,Husa2016,Khan2016}). As a result, the detector strain is formulated as
\be\label{eq:strain}
h(f)= \sqrt{\frac{5}{24}}\frac{  f^{-7/6}}{\pi^{2/3}}\zeta  A_{\rm phen}(f) e^{i[2\pi f t_c + \phi_0 + \Psi_{\rm phen}(f)]}\ ,
\ee
where the termination phase $\phi_0$ \cite{Allen2012} is defined as
\be\label{eq:phase}
e^{2i \phi_0} := e^{2i \phi_c} \frac{F_+(1+\cos^2\iota)/2 -i F_\times \cos\iota}{\sqrt{F_+^2 (\frac{1+\cos^2\iota}{2})^2 + F_\times ^2 \cos^2\iota}}\ ,
\ee
 and the strain amplitude
\be\label{eq:zeta}
\zeta := \frac{\mathcal M_z^{5/6} }{D_{\rm L}}\sqrt{F_+^2 (\frac{1+\cos^2\iota}{2})^2 + F_\times ^2 \cos^2\iota}\ .
\ee

Because of parameter degeneracies, not all the binary parameters can be well constrained even for a loud merger event, e.g.,
the coalescence phase $\phi_c$ is in general weakly constrained due to its degeneracy with angles $\{\iota, \theta, \phi, \psi\}$.
To mitigate the parameter degeneracy, we instead use two detector dependent parameters that are of weak degeneracy with other parameters
(similar parameterization was used in \cite{Roulet2022} for efficient parameter inference):
the strain amplitude $\zeta$ [Eq.~(\ref{eq:zeta})] and the strain phase $\phi_{\rm opt} $ at the optimal frequency
\be\label{eq:phi_opt}
  \phi_{\rm opt} := 2\pi f_{\rm opt} t_c  + \phi_0 + \Psi_{\rm phen}(f_{\rm opt})\ .
\ee
The optimal frequency $f_{\rm opt} := \int \frac{|h(f)|^2 }{P_{\rm n}(f)} f \ df\Big/ \int  \frac{|h(f)|^2 }{P_{\rm n}(f)} \ df$ is the frequency where the waveform is best constrained, with $P_{\rm n}(f)$ being the detector noise PSD.
To summarize, we parametrized the PhenomB waveform with the following 10 model parameters $\{\mathcal M_z, M_z, \chi, \zeta, t_c, \phi_{\rm opt}\}+\{\iota, \theta, \phi, \psi\}$. In this parameterization, the detector strain $h(f)$ depends on the first 6 parameters [see Eqs.~(\ref{eq:strain},\ref{eq:phi_opt})], and the remaining 4 can only be measured with multiple detectors.
Note that $\{\zeta, t_c, \phi_{\rm opt}\}$ are detector dependent quantities and we  choose the most sensitive detector as the reference detector if
multiple detectors are in use. It turns out that this re-parametrization of parameters significantly alleviates the errors from signal subtraction, as discussed later.

\section{Foreground Cleaning Method }
For an incoming binary merger signal $h(f;\bf{\Theta})$ at a detector,
one can infer its ML (Maximum Likelihood) estimate $h(f;{\bf\Theta}^{\rm ML})$ along
with the posterior $\mathcal P({\bf \Theta}|d)$, where $d(f) = h(f) + n(f)$ is
the detector strain data, consisting of signal $h$ and noise $n$.
From the observable $d(f)$, the detector noise PSD $P_{\rm n}(f)$,
and the inferred quantities ${\bf\Theta}^{\rm ML}(d)$ and $\mathcal P({\bf \Theta}|d)$,
one can construct  various foreground cleaning methods.
We shall primarily discuss a method of subtracting the signal from the detector strain and removing
the average residual power in Subsection~\ref{subsec:method1}, then compare it with previous subtraction methods
in Subsection~\ref{subsec:previous}, and discuss an alternative method of directly measuring
the foreground PSD in Subsection~\ref{subsec:method2}.

\subsection{Method 1}\label{subsec:method1}

To clean the foreground, we first
subtract the ML strain $h^{\rm ML}(f)$,
\be\label{eq:sub1a}
\delta h(f) = h(f) - h^{\rm ML}(f)\ .
\ee
Strictly speaking, the correct subtraction should be formulated as $d(f)-h^{\rm ML}(f)$
because the observable is data $d(f)$ instead of signal $h(f)$. But we will use the notation of Eq.~(\ref{eq:sub1a}) for convenience.
After subtracting the ML strain, there is no way to further clean the residual strain $\delta h(f)$,
but the residual power $|\delta h(f)|^2$ is statistically known as
\be\label{eq:deltah2}
  \braket{|\delta h(f;{\bf\Theta})|^2}
  = \braket{\left|h(f;{\bf\Theta})-h(f;{\bf\Theta^{\rm bst}}|{\bf\Theta})\right|^2}\ ,\\
\ee
where $\braket{}$ is the ensemble average over different noise realizations and
${\bf\Theta}^{\rm bst}$ denotes the ML estimate of a signal $h(f;{\bf\Theta})$ in an arbitary noise realization.
For a signal $h(f;{\bf\Theta})$ in a random noise realization, the ML parameter ${\bf\Theta}^{\rm bst}$
can be efficiently pinned down with common optimization algorithms
in the 6-dimensional space  $\{\mathcal M_z, M_z, \chi, \zeta, t_c, \phi_{\rm opt}\}$
when the initial guess is well informed by the posterior $\mathcal{P}({\bf\Theta}|d)$.
The average residual power (\ref{eq:deltah2})   is only known with some uncertainty
informed by the posterior $\mathcal{P}({\bf\Theta}|d)$. Therefore the residual power  estimator
we could construct is
\be
\braket{|\delta h(f;{\bf\Theta}|d)|^2} = \int \mathcal{P}({\bf\Theta}|d)
\braket{|\delta h(f;{\bf\Theta})|^2}
\ d{\bf \Theta}\ ,
\ee
which is  computationally more expensive than Eq.~(\ref{eq:deltah2}) since a high-dimensional integration is involved.
In this work, we will use the approximation $\mathcal{P}({\bf\Theta}|d)\approx \delta({\bf\Theta}-{\bf\Theta}^{\rm ML}(d))$, i.e.,
\be\label{eq:bias}
\braket{|\delta h(f;{\bf\Theta}|d)|^2}\approx \braket{|\delta h(f;{\bf\Theta}={\bf\Theta}^{\rm ML}(d))|^2}\ ,
\ee
which turns out to be a very good approximation inducing a small bias  as we will show
after Eq.~(\ref{eq:var_bias}) and in Figs.~\ref{fig:app1}, \ref{fig:app2}. After removing the average power,
we arrive at the final result,
\be\label{eq:deltaH_rfn}
\delta_1^{\rm rfn} H(f)
=\frac{5}{T}\sum \left\{ |\delta h(f;{\bf\Theta})|^2 -\braket{|\delta h(f;{\bf\Theta}^{\rm ML}(d))|^2} \right\}_i\ , \\
\ee
where the 1st term on R.H.S. is the residual power after subtracting the ML strain,
and the 2nd is the (approximate) average residual power to be removed.

We now  calculate the residual PSD $\delta_1^{\rm rfn} H(f)$.
A simple scaling analysis shows that, $\sigma(\phi_{\rm opt})\sim \sigma(\zeta)/\zeta \sim \rho^{-1}$,
therefore $|\delta h|/|h|\sim \rho^{-1}$ and
the fractional residual power $|\delta h|^2/|h|^2\sim \rho^{-2}$,
where $\rho$ is the signal to noise ratio (SNR).
Considering that $|h|^2\propto \rho^2$, therefore $|\delta h|^2\sim \rho^0$, i.e.,
 the residual power $|\delta h|^2$ is independent from the event SNR.
After removing the average residual power, the fractional residual power is further
reduced by a factor $\sqrt{N_{\rm O}}$ until hitting the bias floor ($N_{\rm O}$ is the total number of mergers detected), i.e.,
\be\label{eq:var_bias}
  \delta_1^{\rm rfn} H(f) =\delta_1^{\rm var} H(f) + \delta_1^{\rm bias} H(f)\ ,
\ee
where
\be
\begin{aligned}
  \delta_1^{\rm var} H(f) &:=\frac{5}{T}\sum^{N_{\rm O}}_{i=1} |\delta h(f;{\bf\Theta})|^2_i -\braket{|\delta h(f;{\bf\Theta})|^2}_i \ , \\
  \delta_1^{\rm bias} H(f)&:=\frac{5}{T}\sum^{N_{\rm O}}_{i=1}\braket{|\delta h(f;{\bf\Theta})|^2}_i-\braket{|\delta h(f;{\bf\Theta}^{\rm ML}(d))|^2}_i\ . \nonumber
\end{aligned}
\ee
Here  $\delta_1^{\rm var} H(f)$ is the variance of a finite number of events,
and  $\delta_1^{\rm bias} H(f)$ is the bias induced by the approximation in Eq.~(\ref{eq:bias}).
The fractional residuals scale with SNR as  $\delta_1^{\rm var}H/H\sim \rho^{-2}N_{\rm O}^{-1/2}$ and $\delta_1^{\rm bias}H/H\sim \rho^{-3}$ {considering that $|\delta h|^2/|h|^2\sim \rho^{-2}$ and
$\delta_1^{\rm bias}H/H\approx
|\delta h|^2_{,\alpha}\delta\Theta^\alpha/|h|^2\sim \rho^{-3}$}, where $\delta{\bf\Theta}={\bf\Theta}-{\bf\Theta}^{\rm ML}$.

Quantitatively, we expand the residual $\delta h$ to $\mathcal O(\rho^{-2})$ as
\be
\delta h = h_{,\alpha}\delta\Theta^\alpha + \frac{1}{2} h_{,\alpha\beta}\delta\Theta^\alpha\delta\Theta^\beta + \mathcal O(\rho^{-3})\ .
\ee
Consequently,
\be
\begin{aligned}
  \sum_i |\delta h(f; {\bf\Theta})|_i^2
  &=  \sum_i |h_{,\alpha}\delta\Theta^\alpha|^2_i +\frac{1}{4}|h_{,\alpha\beta}\delta\Theta^\alpha\delta\Theta^\beta|^2_i\\
  &+ \sum_i |h_{,\alpha}\delta\Theta^\alpha|^2_i \times \mathcal O\left(\frac{\rho^{-1}}{\sqrt{N_{\rm O}}}\right)_i\ , \\
  \braket{|\delta h(f; {\bf\Theta})|^2 }
  =& C^{\alpha\beta}({\bf\Theta})h_{,\alpha}^\star h_{,\beta}\\
  +& \frac{1}{4}h_{,\alpha\beta}^\star h_{,\gamma\tau}
  (C^{\alpha\beta}C^{\gamma\tau} +C^{\alpha\gamma}C^{\beta\tau} +C^{\alpha\tau}C^{\gamma\beta} )\ ,
\end{aligned}
\ee
and similarly for $\braket{|\delta h(f; {\bf\Theta}^{\rm ML}(d))|^2 }$,
where we have used the fact $\braket{\delta\Theta^\alpha\delta\Theta^\beta\delta\Theta^\gamma}=0$
and $C^{\alpha\beta}({\bf\Theta}):=\braket{\delta\Theta^\alpha\delta\Theta^\beta}$ is the covariance matrix.
Plugging the above equations into Eq.~(\ref{eq:var_bias}), we obtain
\be\label{deltaH_rfn_lin}
\begin{aligned}
\delta_1^{\rm rfn} H(f)
  &=\frac{5}{T} \sum_i \left\{  |h_{,\alpha}\delta\Theta^\alpha|^2
  -C^{\alpha\beta}({\bf\Theta}^{\rm ML}(d)) h^{\star{\rm ML}}_{,\alpha} h^{\rm ML}_{,\beta} \right\}_i \\
  &+\frac{5}{T} \sum_i  \left\{|h_{,\alpha}\delta\Theta^\alpha|^2\times \left[\frac{\mathcal O(\rho^{-1})}{\sqrt{N_{\rm O}}}
  + \mathcal O(\rho^{-3}) \right]\right\}_i\ ,
\end{aligned}
\ee
where the 2nd row on the R.H.S. contributes as a small correction to the 1st row.
As a conservative estimate, we will take  $O(\rho^{-1})=10\rho^{-1}$ and $O(\rho^{-3})=10\rho^{-3}$ in the following calculation.
For reference, we denote the residual PSD after subtracting the ML strain and before removing the average power as
\be\label{deltaH_lin}
\delta_1  H(f) = \frac{5}{T} \sum_i   |h_{,\alpha}\delta\Theta^\alpha|^2_i \ .
\ee

In summary, our method has achieved a two-step noise reduction: using a new set of binary parameters for event subtraction to obtain  $\delta_1  H(f)$ and performing a further residual power subtraction to arrive at  $\delta_1^{\rm rfn} H(f)$.

\subsection{Comparison with previous subtraction methods}\label{subsec:previous}

At this stage, it is informative to compare the previous subtraction methods \cite{Sachdev2020,Cutler2006}. In the state of the art work by Regimbau, Sachdev and
Sathyaprakash \cite{Sachdev2020},
the residual PSD after subtraction is formulated as
\be\label{eq:soa}
\delta H(f) =
\frac{1}{T}\sum_i  \left(|h_+(f)-h_+^{\rm ML}(f)|^2+|h_\times(f)-h_\times^{\rm ML}(f)|^2 \right)_i \ .
\ee
The first subtlety of this subtraction method is that it is difficult to be applied to real data, because
the observable is the detector strain $d$ instead of the two polarizations $h_{+,\times}(f)$ in Eq.~(\ref{eq:soa}). The second issue, which is somewhat related to the first one and as noticed in Refs.~\cite{Zhou2022,Zhou2022b}, is the resulting high fractional residue, with
 $\delta H(f)/H(f)\sim 50\%$ even for BBHs detected with 3G detectors.
This residual level is much higher than $\delta_1  H(f)/H(f)$ with the $\rho^{-2}$ scaling (c.f. Eq.~\ref{deltaH_lin}),
simply because
\be
|h_{+,\times}(f)-h_{+,\times}^{\rm ML}(f)|\approx |h_{+,\times}(f)(e^{i\phi_c}-e^{i\phi_c^{\rm ML}})|=
|h_{+,\times}(f)|\times|1-e^{i\delta\phi_c}|
\ee
and the coalescence phase $\phi_c$ is weakly constrained with uncertainty $\sigma(\phi_c)=\mathcal O (1)$ due to the parameter degeneracy (see Fig.~7 in \cite{Zhou2022} for detailed numerical analysis). Notice that the actual observed signal $h$ from data $d$ is insensitive to this parameter degeneracy that affects $h_{+,\times}$, so it is natural to work with $h$ instead of $h_{+,\times}$.

In our subtraction method, a different parameterization is used where the parameter degeneracy is largely mitigated with $\sigma(\phi_{\rm opt})\sim \sigma(\zeta)/\zeta \sim \rho^{-1}$ \cite{Roulet2022},
and the subtraction is performed on the detector strain $d$ instead of the polarizations $h_{+,\times}(f)$.
As a result,  a much better precision $\delta_1 H(f)/H(f)\sim |\delta h|^2/|h|^2\sim (\delta\zeta)^2/\zeta^2+(\delta\phi_{\rm opt})^2 \sim \rho^{-2}$ is achieved.

In the original work \cite{Sachdev2020}, the authors assumed that for any BBH signals,
7 of the waveform parameters ${\mathbf\Theta}$ are known except \{$t_c, \phi_c, \mathcal M_z$\}.
With this simplified but less realistic assumption, the degeneracy of the coalescence phase $\phi_c$ with other parameters
is broken and its uncertainty is strongly suppressed with $\sigma(\phi_c)\sim \rho^{-1}$.
As a result, they found a fractional residue $\delta H(f)/H(f)\sim \rho^{-2}$,
which has been shown to be an artifact of incorrectly fixing binary parameters \cite{Zhou2022,Zhou2022b}.

\vspace{0.2 cm}

In another well known work on foreground cleaning method \cite{Cutler2006,Harms2008}, the residual strain $\delta h$ was proposed to
be further cleaned by removing its projection along the tangential space
$\ket{h^{\rm ML}_{,\alpha}}\bra{h^{\rm ML}_{,\beta}}$, i.e.,
\be
\delta h^\perp =\delta h - \delta h^\parallel = \delta h - F^{-1}_{\alpha\beta} \ket{h^{\rm ML}_{,\alpha}}\braket{h^{\rm ML}_{,\beta} |\delta h}
\ee
where the inner product is defined as
\be
\braket{h|g}=  4 \int_0^\infty \frac{{\rm Real} \{h^\star(f) g(f)\} }{P_{ {\rm n}}(f)} df\ ,
\ee
with $P_{ {\rm n}}(f)$ being the detector noise PSD,
and the Fisher matrix is defined as
$F_{\alpha\beta} = \braket{h_{,\alpha}|h_{,\beta} }$.
Expanding the residual  strain as $\delta h=h_{,\alpha} \delta \Theta_\alpha +
\mathcal O( (\delta\Theta)^2)$,
it is straightforward to see that the linear deviation part in $\delta h$  is removed
and the fractional residue scales as $|\delta h^\perp|^2/|h|^2 \sim \rho^{-4}$,
if the above procedure worked out as claimed in \cite{Cutler2006,Harms2008}.

This is in fact not achievable. If this were achieved, it means {\it for a generic single event}, one could measure the signal with error well below the detector noise level,
i.e., the measurement accuracy were not limited by the detector noise, which is counter-intuitive.
In a more quantitative way,  the reason is that the residual data $\delta h+ n$ is known while $\delta h$ is not,
and the projection $ \ket{h^{\rm ML}_{,\alpha}}\braket{h^{\rm ML}_{,\beta}|\delta h+n}$ vanishes exactly, because
the ML strain is defined such that $\braket{d-h^{\rm ML}|d-h^{\rm ML}}$ minimizes, which gives $\braket{h^{\rm ML}_{,\alpha}|d-h^{\rm ML}}
=\braket{h^{\rm ML}_{,\alpha}|\delta h+ n}=0$. This projection and removal procedure works only if
the detector noise vanishes or
some approximate ML strain $h^{\rm prox}$
instead of the ML strain  is subtracted with $\delta h = h-h^{\rm prox}\neq h-h^{\rm ML}$.
For example, in the case of multiple detectors as considered in \cite{Sharma2020}, the ML strain $h^{\rm ML}$ of multi-detector outputs will be
different from the ML strain of each detector output $h^{\rm ML}_{(k)}$, and  the inner product $\braket{h^{\rm ML}_{,\alpha}|d_{(k)}-h^{\rm ML}}$ does not vanish.
In this setting up, the authors of \cite{Sharma2020} found the projection and removal procedure
improves the fractional residue  to the advertised level $|\delta h^\perp_{(k)}|^2/|h_{(k)}|^2 \sim \rho^{-4}$ if there was no detector noise,
while in the presence of detector noise, it is of no surprise to find that this procedure restores the fractional residual $\rho^{-2}$,
which is well above the initially claimed  $\rho^{-4}$ level (see the blue v.s. orange curves in the bottom panel of Fig.~8 in \cite{Sharma2020}).

The numerical results  in \cite{Sharma2020} is straightforward to understand simply because there is no way to evade the detector noise limit
and measure a single signal with uncertainty better than $|\delta h|^2/|h|^2\sim \rho^{-2}$.
This general conclusion can be explicitly shown with a simple likelihood analysis for this specific case.
Starting with a single detector with detector noise PSDs $P_{\rm n}(f)$, the likelihood $L(d|{\bf\Theta})$
of seeing data $d$ given a waveform model with parameter ${\bf\Theta}$ is defined as
\be
\log L(d|{\bf\Theta}) = -\frac{1}{2}\braket{d-h({\bf\Theta})|d-h({\bf\Theta})}\ .
\ee
Considering a small perturbation from the true parameters ${\bf\Theta} = {\bf\Theta}^{\rm true}+\delta{\bf\Theta}$, the waveform expands to the linear order as
 $h({\bf\Theta}) = h({\bf\Theta}^{\rm true}) + h_{,\alpha}\delta\Theta^\alpha$.
 Consequently, the likelihood can be formulated as
 \be
-2 \log L(d|{\bf\Theta}) = (\delta\Theta^\alpha-\delta\Theta^\alpha_{\rm noise}) F_{\alpha\beta} (\delta\Theta^\beta-\delta\Theta^\beta_{\rm noise})\ ,
 \ee
 with  $\delta\Theta^\alpha_{\rm noise} = (F^{-1})^{\alpha\beta} \braket{h_{,\beta}|n}$ being the parameter shift driven by the detector noise
 (see e.g. \cite{Flanagan1998,Cutler2007,Antonelli2021} for details). The best-fit or the ML parameters are therefore ${\bf\Theta}^{\rm ML}={\bf\Theta}^{\rm true}+\delta{\bf\Theta}^{\rm noise}$.
 Going back to the multiple-detector case, the joint likelihood is therefore
 \be
-2 \log L(d|{\bf\Theta}) = \sum_k (\delta\Theta^\alpha-\delta\Theta^\alpha_{(k)\rm noise}) F_{(k)\alpha\beta} (\delta\Theta^\beta-\delta\Theta^\beta_{(k)\rm noise})\ ,
 \ee
 and the ML parameters are determined by maximizing the joint likelihood, i.e.,
 \be
 0  =  \sum_k F_{(k)\alpha\beta} (\delta\Theta^\beta-\delta\Theta^\beta_{(k)\rm noise})\ ,
 \ee
 where $F_{(k)\alpha\beta}:= \braket{h_{(k),\alpha}|h_{(k),\beta}}$ is the Fisher matrix of detector $k$.
 Considering a simple case of two detectors with a same noise PSD and a same orientation, i.e., $F_{(1)} = F_{(2)}$, the ML parameters are
 ${\bf\Theta}^{\rm ML}={\bf\Theta}^{\rm true}+\frac{1}{2}(\delta{\bf\Theta}^{\rm noise}_{(1)}+\delta{\bf\Theta}^{\rm noise}_{(2)})$.
Applying the projection and removal procedure to the detector 1 data, we obtain the residual strain
 \be\label{eq:delta_h_perp}
 \begin{aligned}
 \delta h^\perp_{(1)}
 &= \delta h_{(1)} - F^{-1}_{(1)\alpha\beta} \ket{h^{\rm ML}_{,\alpha}}\braket{h^{\rm ML}_{,\beta} |\delta h_{(1)} + n_{(1)}}\ , \\
 &= \delta h_{(1),\alpha}\left(\frac{3}{2}\delta\Theta^{\alpha}_{(1){\rm noise}} -\frac{1}{2}\delta\Theta^\alpha_{(2){\rm noise}}  \right) + \mathcal O(\delta\Theta)^2\ .
\end{aligned}
 \ee
Consequently, the  fractional residual PSD is naturally $| \delta h^\perp_{(1)}|^2/|h_{(1)}|^2\sim \rho^{-2}$ as numerically confirmed in \cite{Sharma2020},
instead of $\rho^{-4}$ as claimed in \cite{Cutler2006,Harms2008}. It is straightforward to
generalize the derivation of Eq.~(\ref{eq:delta_h_perp}) to multiple-detector cases.

\vspace{0.2 cm}

In  summary, the astrophysical foreground can be cleaned with fractional residue $\delta H/H\sim \rho^{-2}$
with the state of the art subtraction method proposed by Regimbau, Sachdev and Sathyaprakash \cite{Sachdev2020}
if the binary model parameters $\mathbf{\Theta}$ were known except $\{t_c, \phi_c, \mathcal M_z\}$, while the fractional
residue turns out to be $\sim \mathcal O(1)$ without fixing any model parameters a priori as shown in recent papers \cite{Zhou2022,Zhou2022b}.
Another possibly more serious issue  is that it is unclear how to apply it to real data, because this method is designed to apply to
GW polarizations $h_{+,\times}$ instead of detector strain $d$.
With the projection method proposed in  \cite{Cutler2006,Harms2008},
the astrophysical foreground can be cleaned with fractional residue $\delta H/H\sim \rho^{-4}$  if there were no detector noise, while the fractional residue turns out to be $\sim \rho^{-2}$ in the presence of detector noise \cite{Sharma2020}. As we will show in the next section,
the astrophysical foreground can be cleaned with fractional residue $\delta H/H\sim \rho^{-3}$ as we consider a family of events with our method.

\subsection{Method 2}\label{subsec:method2}

 The basic picture of Method 1
 is subtracting the unknown signal $h(f)$ from data $d(f)$
with the model $h^{\rm ML}(f)$ as a proxy, where the precision of strain phase measurement
makes a big difference. As a result, the residual $\delta_1 H(f)$ or $\delta_1^{\rm rfn}H(f)$
is in general lowest around the optimal frequency $f_{\rm opt}$
where the phase of the detector strain is  best measured,
and the fractional residual blows up at much lower or much higher frequencies where the phase is not well constrained (see Fig.~\ref{fig:app1}).
On the other hand, the foreground energy density or the PSD depends only on the strain amplitude
$\Omega_{\rm GW}(f)\propto H(f)\propto\sum_i |h(f)|_i^2 $. Therefore it is possible to
measure the PSD using the amplitude information only.

For this purpose, an obvious estimator to use would be
\be
\hat H(f) = \frac{5}{T}\sum_i  |h^{\rm ML}(f)|^2_i\ ,
\ee
which is in fact a biased estimator with  $H(f)-\braket{\hat H(f)} < 0$.
The above primitive estimator can be refined by compensating the bias as
\be
\hat H^{\rm rfn}(f) = \frac{5}{T}\sum_i  \left(|h^{\rm ML}(f)|^2-\braket{|\sigma_h(f)|^2}\right)_i\ ,
\ee
where $-\braket{|\sigma_h(f;{\bf\Theta})|^2} := |h(f;{\bf\Theta})|^2-\braket{|h(f; {\bf\Theta}^{\rm bst}|{\bf\Theta})|^2}$
is the compensation term. It can be computed if the true parameters ${\bf \Theta}$ were known, while ${\bf \Theta}$ is only known with some uncertainty informed by the posterior $\mathcal P({\bf\Theta}|d)$.
Using the same approximation $\mathcal P({\bf\Theta}|d)\approx \delta({\bf\Theta}-{\bf\Theta}^{\rm ML}(d))$
as in Method 1, we have
\be
\braket{|\sigma_h(f;{\bf\Theta})|^2} \approx \braket{|\sigma_h(f;{\bf\Theta}={\bf\Theta}^{\rm ML}(d))|^2}\ .
\ee
As a result,  the final form of the refined estimator is
\be
\hat H^{\rm rfn}(f) = \frac{5}{T}\sum_i  \left(|h^{\rm ML}(f)|^2- \braket{|\sigma_h(f;{\bf\Theta}^{\rm ML}(d))|^2}\right)_i\ ,
\ee
with residual PSD
\be
\delta_2^{\rm rfn} H(f) = \left|H(f)-\hat H^{\rm rfn}(f)\right|\ .
\ee

Similar to in Method 1,  the residual PSD can be computed with the noise PSD $P_{\rm n}(f)$.
Making use of  the fact that the uncertainty in amplitude
$|h(f)|=\zeta A_{\rm phen}(f;\mathcal M_z, M_z, \chi)$ is mainly sourced
by the uncertainty in strain amplitude parameter $\zeta$,
the bias term can be further approximated as
\be\label{eq:sigma_zeta}
 \braket{ |\sigma_h(f;{\bf\Theta}^{\rm ML}(d))|^2 }
\approx \left|\frac{\sigma(\zeta)}{\zeta^{\rm ML}} h^{\rm ML}(f)\right|^2\ ,
\ee
where $\zeta^{\rm ML}$ is the ML strain amplitude and $\sigma(\zeta)=\sqrt{C^{\zeta\zeta}({\bf\Theta}^{\rm ML})}$ is its 1-$\sigma$ uncertainty.
With this approximation, we obtain a conservative estimate of the residual PSD
\be\label{eq:delta2_H_rfn}
\delta_2^{\rm rfn} H(f) \approx
\frac{5}{T}\left| \sum_i  |h(f)|_i^2- |h^{\rm ML}(f)|_i^2+ \left|\frac{\sigma(\zeta)}{\zeta^{\rm ML}} h^{\rm ML}(f)\right|^2_i \right| \ .
\ee
For comparison use, we denote the residual PSD of the primitive estimator as
\be\label{eq:delta2}
  \delta_2 H(f) =\left| H(f)-\hat H(f) \right|\ .
\ee

\section{Foreground Cleaning With 3G Detectors}

\begin{figure}
\includegraphics[scale=0.7]{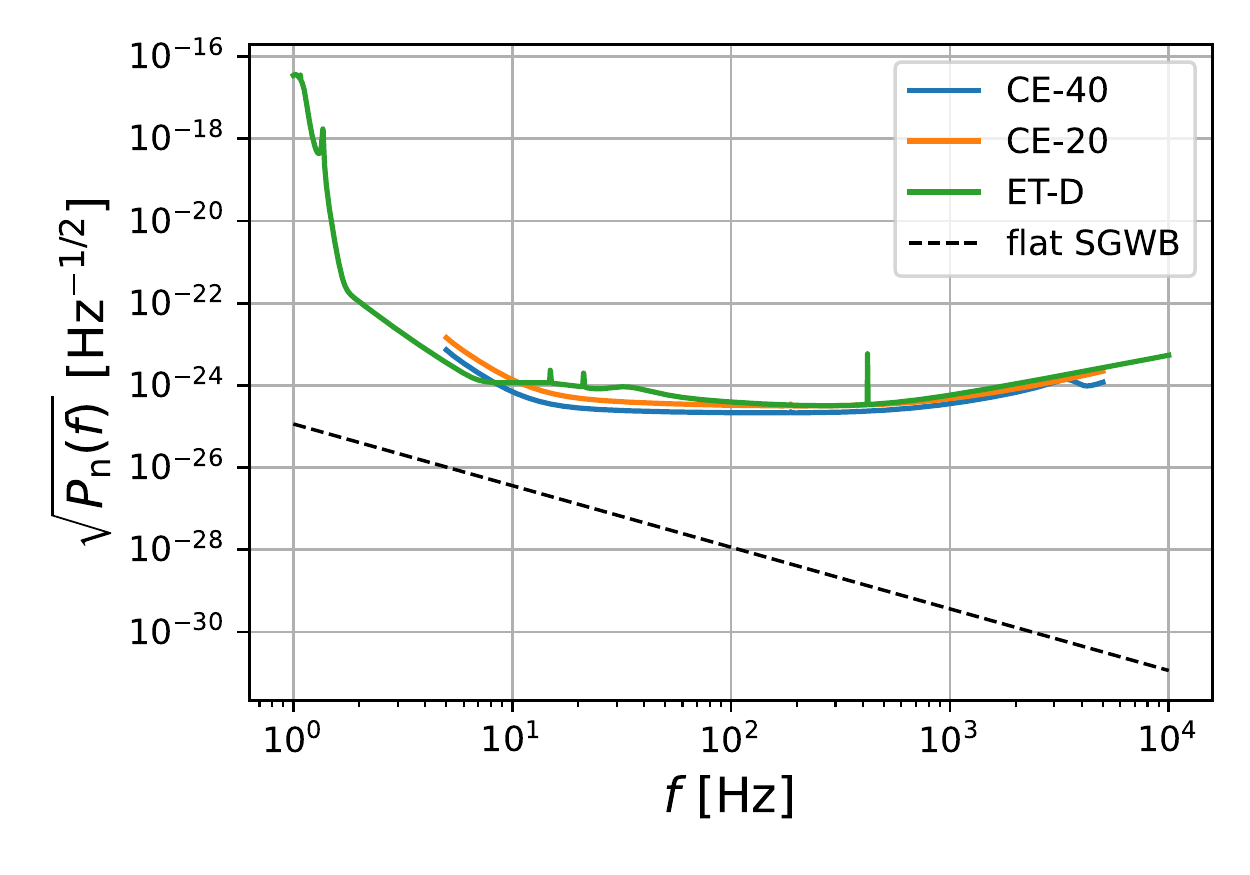}
\caption{\label{fig:Pnf}
Noise PSD $P_{\rm n}(f)$ of 3G detectors considered. For comparison, we also plot the noise PSD sourced by
a flat-spectrum SGWB with energy density $\Omega_{\rm SGWB}(f)=10^{-13}$ in the dashed line.}
\end{figure}

We now consider a population of BBH/BNS mergers and
apply the foreground cleaning methods to a mock observation of 3G detectors.
Following the discussion in Ref.~\cite{Borhanian2021}, we consider a 3G detector network:  CE\_40 (Idaho, USA) + CE\_20 (New South Wales, Australia) + ET\_D (Cascina, Italy)
consisting of a stage-2 40-km compact-binary optimized CE, a stage-2 20-km compact-binary optimized CE, and an ET of type D (see \cite{Maggiore2020,Srivastava2022,Borhanian2021} for details about detector sensitivities, locations and orientations).  The detector noise PSDs are plotted in Fig.~\ref{fig:Pnf}. As a reference model in this paper,
 we consider  a flat-spectrum SGWB with energy density $\Omega_{\rm SGWB}(f)=10^{-13}$.
For comparison, we also plot the noise PSD $P_{\rm n, SGWB}(f)$  sourced by this background,
with $P_{\rm n, SGWB}(f) = \frac{1}{5}\frac{3H_0^2}{4\pi^2} f^{-3} \Omega_{\rm SGWB}(f)$ [see Eq.~(\ref{eq:omega})].

\subsection{Cleaning the BBH Foreground with Method 1}

In a population model, we need to specify the volumetric merger rate $R(z)$ (number of mergers per comoving volume per cosmic time at redshift $z$),
the mass distribution $p(m_1,m_2)$ and the effective spin distribution $p(\chi)$.
The merger rate in the observer frame is written as
\be
\dot N = \int \frac{dV_{\rm c}(z)}{dz}\frac{R(z)}{1+z} \ dz \ ,
\ee
where $V_{\rm c}(z)$ is the comoving volume up to redshift $z$, and the factor $1+z$ comes from the time dialation due to cosmic expansion.
Consistent with the LVK O1-O3 observations \cite{LVCO1O2,LVKCO3}, we take the BBH  merger rate as
$R_{\rm BBH}(z) = R_0\times (1+z)^{2.9} e^{-z^2/3}$ for ($z\leq 6$),
with the local merger rate $R_0=20 \ {\rm Gpc}^{-3}{\rm yr}^{-1}$ (see e.g. \cite{Perigois2021,Zhou2022,Regimbau2022}
for more detailed rate modeling), a spin distribution
$p(\chi)$ as a Gaussian distribution with a mean value $0.06$ and a standard deviation $0.1$,
and a mass  distribution
$p(m_1, m_2) \propto m_1^{-1} (m_1-m_{\rm min})^{-1}$ for $m_{\rm min}\leq m_1 \leq m_2 \leq m_{\rm max}$,
with $m_{\rm min} = 5 M_\odot$ and $m_{\rm max} = 42 M_\odot$.
In this population model, the total BBH merger rate turns out to be $\dot N_{\rm BBH} = 4.6\times 10^4\ {\rm yr}^{-1}$,
and the BBH foreground energy density is $\Omega_{\rm gw}(f)\approx 0.8\times 10^{-10}\times (f/{\rm Hz})^{2/3}$ for $f\lesssim 200$ Hz.

\begin{figure}
\includegraphics[scale=0.7]{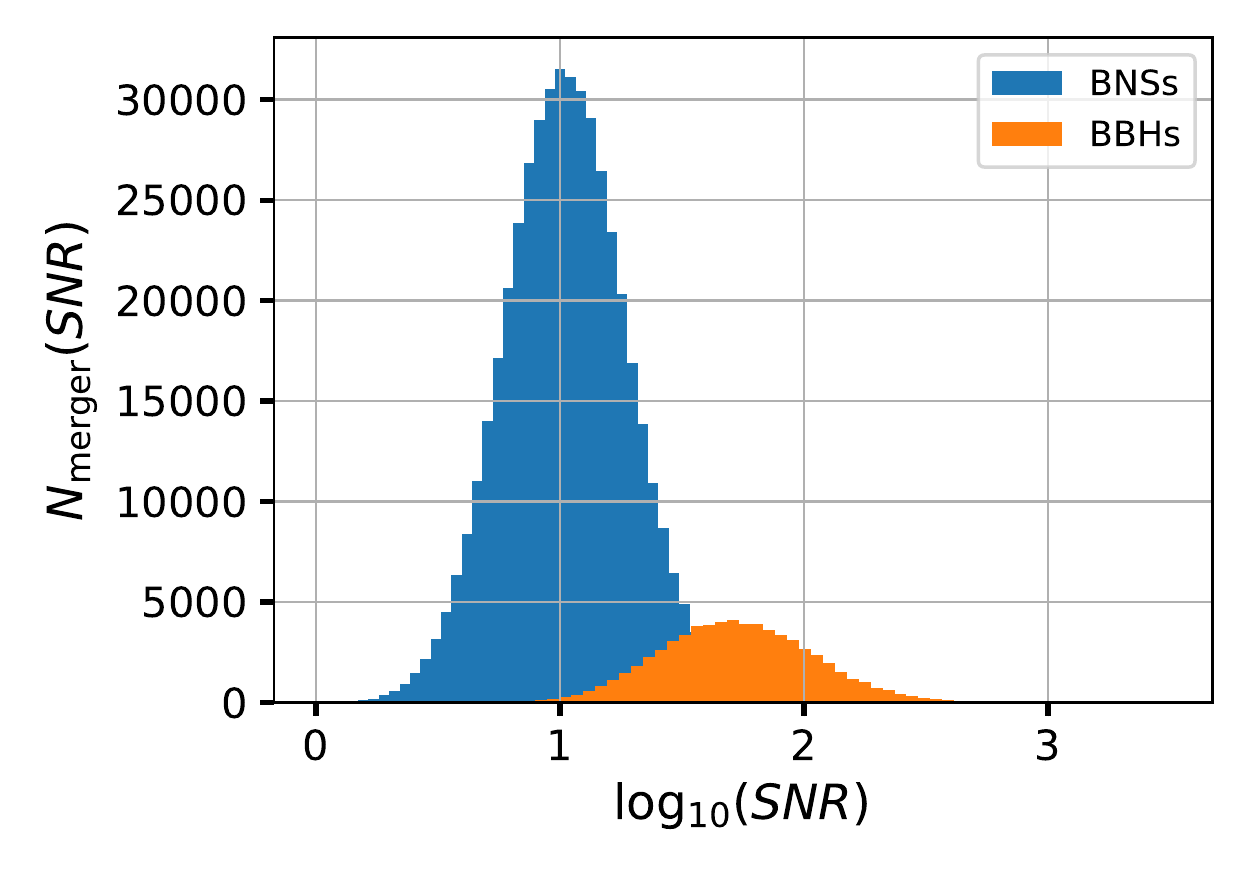}
\caption{\label{fig:N_snr}
SNR historgrams of a BBH population realization and a BNS population realization we used.}
\end{figure}

We generate 16 BBH population realizations, with $6.5\times10^4$ BBH mergers in each realization (that corresponds to approximately 1.4 years of observation with the assumed BBH merger rate).
The BH masses, spins, and redshifts are sampled according to the distributions specified above,
and all the angles are sampled assuming isotropy. For each merger, we calculate the expected SNR
as $\rho = \sqrt{\sum_{ {\rm det.}\ k}\braket{h_{(k)}|h_{(k)}}}$, where the inner product is defined as
\be
\braket{h_{(k)}|g_{(k)}}= 4 \int_0^\infty \frac{{\rm Real} \{h^\star_{(k)}(f) g_{(k)}(f)\} }{P_{ {\rm n}, (k)}(f)} df\ ,
\ee
with $P_{{\rm n},(k)}(f)$ and $h_{(k)}$ being the noise PSD and the signal strain
of the $i$-th detector, respectively.
We find the merger SNR distribution peaks around $30$ with a long tail extending to several hundred, and almost all the merger are of SNR $>10$
 (see Fig.~\ref{fig:N_snr}).
For latter parameter inference use, we also calculate the Fisher information matrix as
$F_{\alpha\beta} = \sum_k F_{(k)\alpha\beta} = \sum_{ {\rm det.}\ k}\braket{h_{(k),\alpha}|h_{(k),\beta}} $.

\begin{figure}
\includegraphics[scale=0.6]{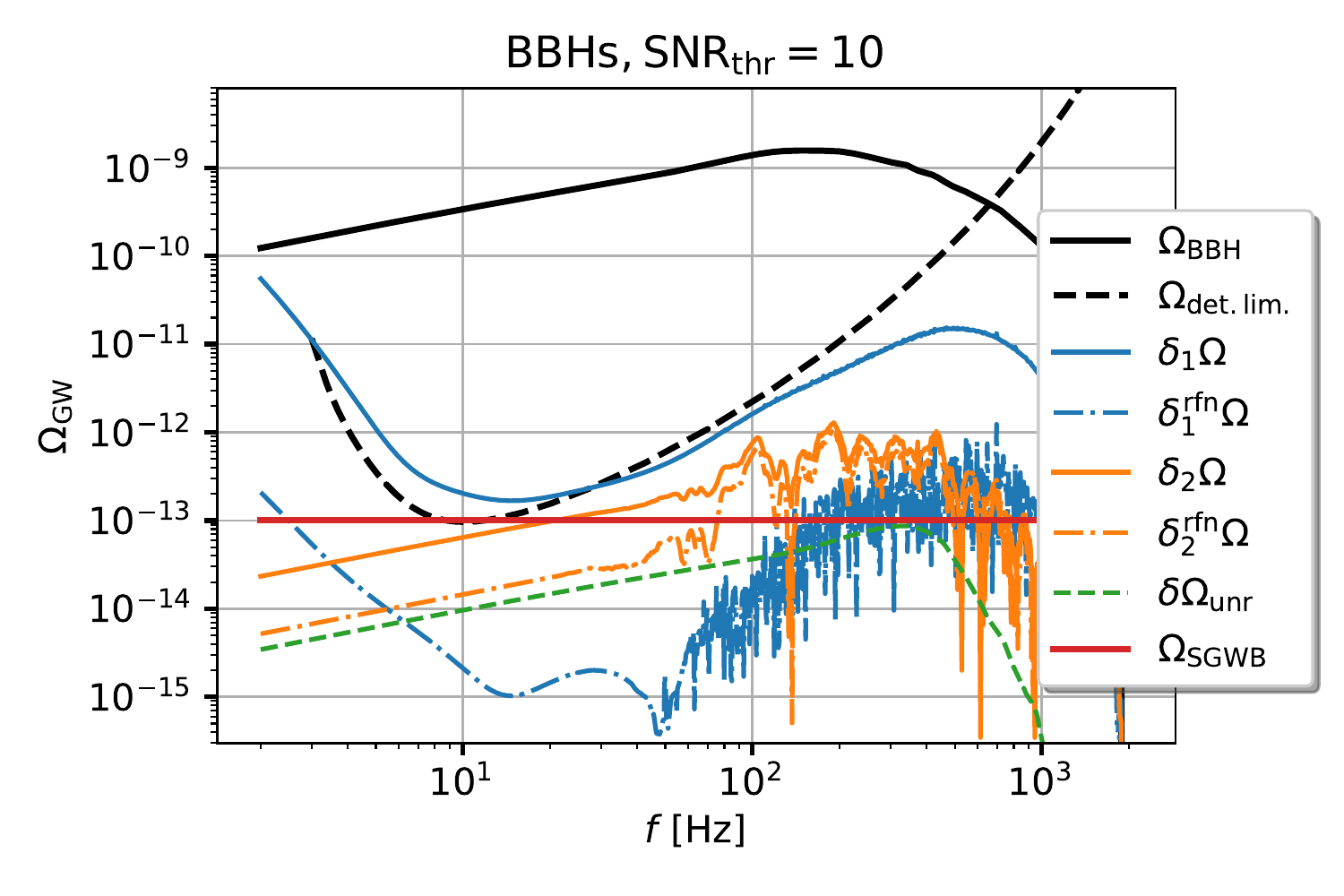}
\caption{\label{fig:app1} Residuals after cleaning the BBH foreground.
The top black solid/dashed lines are the total energy density $\Omega_{\rm GW}$ of the BBH foreground and
the detector sensitivity limit $\Omega_{\rm det.lim.}$, respectively.
The blue solid/dot-dashed lines are the energy density of the residual foreground after implementing
the primitive [$\delta_1\Omega$, Eq.~(\ref{deltaH_lin})] and refined [$\delta_1^{\rm rfn}\Omega$, Eq.~(\ref{deltaH_rfn_lin})] subtractions in Method 1, respectively.  Here the residue $\delta_1^{\rm rfn}\Omega$
is dominated by the bias term in Eq.~(\ref{eq:var_bias}).
The orange solid/dashed lines are the energy density of the residual foreground after implementing
the primitive [$\delta_2\Omega$, Eq.~(\ref{eq:delta2})] and refined [$\delta_2^{\rm rfn}\Omega$, Eq.(\ref{eq:delta2_H_rfn})] estimates in Method 2, respectively.
The green dashed line is the energy density $\Omega_{\rm unr}$ of GWs from unresolved BBH mergers with $\rho< 10$.
 The horizontal red solid line is the energy density $\Omega_{\rm SGWB}$ of the reference flat SGWB.}
\end{figure}

For each merger with model parameter $\bf\Theta$, we sample its ML parameters
$\bf\Theta^{\rm ML}$ from a multivariate Gaussian distribution with a mean value $\bf\Theta$ and
a covariance matrix ${\bf C(\Theta)} :={\bf F}^{-1}$,   which simulates the effect of
Gaussian noise in a ML search.  We process the mergers with $\rho > \rho_{\rm thr}=10$ through the foreground cleaning processes with the most sensitive
 CE\_40 as the reference detector, and label the remaining mergers as unresolved, i.e.,
 \be
\Omega_{\rm res} = \delta\Omega + \Omega_{\rm unr}\ ,
 \ee
 where $\Omega_{\rm res}$ is the total residue,
 $\Omega_{\rm unr}$ is the energy density of foreground from unresolved mergers
 and $\delta\Omega $ is the residue of cleaning the resolved mergers.
 We average all different residuals over the 16 realizations simulated (see Fig.~\ref{fig:app1}). The black dashed line is the detector sensitivity limit $\Omega_{\rm det.lim.}(f)$,  which is the sensitivity of the detector network to the SGWB if there were no astrophysical foreground (a commonly used definition is the power-law integrated sensitivity curve proposed in \cite{Thrane2013}). Quantitatively, it
 is defined such that any SGWB with energy density $\Omega_{\rm SGWB}(f)$ that is tangent to the detector sensitivity limit curve at $f_0$, i.e.,
\be
\Omega_{\rm SGWB}(f)= \Omega_{\rm det.lim.}(f_0)\times(f/f_0)^{\gamma_0}
\ee
with the power index $\gamma_0 = \frac{d\ln \Omega_{\rm det.lim.}(f_0)}{d\ln f_0}$,
can be detected by the detector network with $3\sigma$ confidence level in 4 years
if there was no foreground contamination (\cite{Moore2015,Zhou2022} for the computational details).

After subtracting the ML strain, the fractional residual PSD [Eq.~(\ref{deltaH_lin})] turns out to be
$\delta_1 H/H=\delta_1\Omega/\Omega_{\rm GW}\approx 3\times 10^{-4}\sim N_{\rm O}/\sum_i\rho_i^{2}$ at $f\approx 20$ Hz.
After removing the average residual power [Eq.~(\ref{deltaH_rfn_lin})],
the fractional residual PSD further improves to $\approx 3\times 10^{-6}\sim \sum_i\rho_i^{-1}/\sum_i\rho_i^2$
at the same frequency, which shows that the residual is dominated by the small bias  $\delta_1^{\rm bias} H(f)$
induced by the approximation in Eq.~(\ref{eq:bias}).
And the residual energy density $\delta_1^{\rm rfn}\Omega$ turns out to be
below the detector sensitivity $\Omega_{\rm det. lim.}$  by a factor of $\mathcal O(10^2)$
in  the whole frequency range.
In addition, the energy density $\Omega_{\rm unr}$ of unresolved BBH foreground
is always  below the detector sensitivity limit $\Omega_{\rm det.lim.}$  by a factor of $\mathcal O(10)$.

Note the simplified simulations and subsequent
ML parameter sampling process used in this work are not entirely realistic.
In a more realistic simulation, the time series of detector strain should be simulated as the summation of merger signals and detector noise, and the ML parameters should be
inferred from the simulated data by an optimization algorithm. For 3G detectors, the overlapping signals make this parameter inference process more complicate considering the abundance of merger signals, where the inference of
parameters of multiple signals and the number of signals should be performed simultaneously.
Though recent studies show that overlapping signals will produce serious biases in the parameter inference
in rare cases (less than 1 occurrence per year for 3G detectors) where the coalescence time and the chirp masses of the two overlapping signals are very close to each other  \cite{Himemoto2021,Pizzati2022},
the impact on the foreground cleaning problem remains to be explored.
In addition, the Fisher analysis may predict lower parameter uncertainties by a factor of $\mathcal O(1)$ for
low-SNR events, thus the foreground residue estimation based on the Fisher matrix approach may be lower by a factor of $\mathcal O(1)$.
In this work, we limit our investigation to the simplified simulation as a proof-of-principle
for the foreground cleaning methods.

\subsection{Cleaning the BNS Foreground with Method 1 } Similar to the BBH population,
we take the BNS  merger rate as $R_{\rm BNS}(z) = R_0\times (1+z)^{2.9} e^{-z^2/3}$ for ($z\leq 6$),
with the local merger rate $R_0=160 \ {\rm Gpc}^{-3}{\rm yr}^{-1}$, a spin distribution
$p(\chi)$ as a Gaussian distribution with a mean value $0.03$ and a standard deviation $0.03$.
and a uniform mass distribution between $1.1 M_\odot$ and $2.1 M_\odot$.
In this population model, the total merger rate turns out to be $\dot N_{\rm BNS} = 3.7\times 10^5\ {\rm yr}^{-1}$,
and the GW foreground energy density is $\Omega_{\rm gw}(f)\approx 1.5\times 10^{-11}\times (f/{\rm Hz})^{2/3}$ for $f\lesssim 2000$ Hz.

\begin{figure}
\includegraphics[scale=0.6]{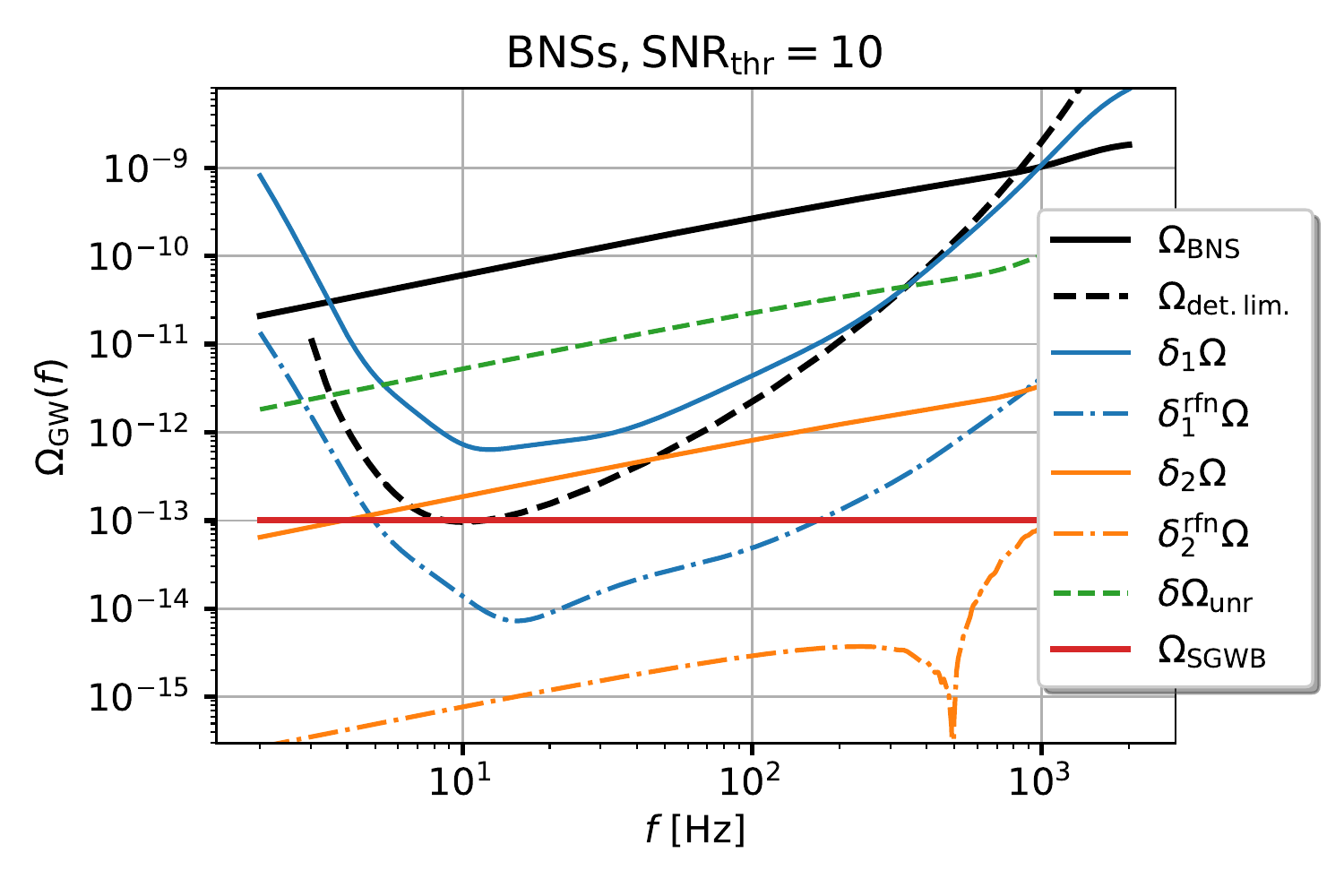}
\caption{\label{fig:app2} Same to Fig.~\ref{fig:app1} except for BNSs.}
\end{figure}

We generate 16 BNS population realizations, with $5.2\times10^5$ BNS mergers in each realization (roughly 1.4 years of observation).
We find the merger SNR distribution peaks around $8$ and about half of the BNSs are sub-threshold with $\rho < \rho_{\rm thr}=10$.
Implementing the foreground cleaning method above,
the fraction residual PSD of BNSs with $\rho > \rho_{\rm thr}$ turns out to be
$\delta_1^{\rm rfn} H/H\approx  10^{-4}\sim \sum_i\rho_i^{-1}/\sum_i\rho_i^{2}$ at $f\approx 20$ Hz,
and the residual energy density $\delta^{\rm rfn}_1\Omega$ is well below the detector sensitivity limit $\Omega_{\rm det.lim.}$ in the whole frequency range.
Due to a large fraction of low-SNR BNSs in the population,
the sub-threshold BNSs turns out to dominate the residual foreground
with  $\Omega_{\rm unr}(f)\approx 1.4\times 10^{-12}\times (f/{\rm Hz})^{2/3}$,
which is well above  the detector
sensitivity limit in a large frequency range (similar conclusion had also been reached in previous works \cite{Regimbau2017,Sachdev2020,Zhong2022}).

\subsection{Foreground cleaning with Method 2}

We also apply the Method 2 to the simulated BBHs and BNSs in the maintext,
and compare the preformance of the two methods in
Figs.~\ref{fig:app1} and \ref{fig:app2}.
With the primitive estimate in Method 2 [Eq.~(\ref{eq:delta2})], the residual energy density $\delta_2\Omega$ of BBHs is already below $\Omega_{\rm det.lim.}$
(with the fractional residual $\delta_2\Omega/\Omega_{\rm GW}\approx 2\times 10^{-4}\sim N_{\rm O}/\sum_i\rho_i^{2}$ ) across the whole frequency range
and the refined estimate [Eq.(\ref{eq:delta2_H_rfn})] further improves the residual by a factor $\sim 4$ at low frequency $f\lesssim 10^2$ Hz. The improvement factor is much lower than $\sqrt{N_{\rm O}}$ because the bias
term of each merger is of different magnitude with $|\sigma_h(f;{\bf\Theta}^{\rm ML}(d))|^2_i\propto \rho_i^{-2}$, therefore the fractional residual decreases slower than the scaling $N_{\rm O}^{-1/2}$. Applying the primitive estimate in Method 2 to the BNSs that are individually detectable, we find
the fractional residual energy density $\delta_2\Omega/\Omega_{\rm GW}\approx 3\times 10^{-3}\sim N_{\rm O}/\sum_i\rho_i^2$ and  the refined estimate further improves the residual by a factor $\sim 3\times 10^2$, which is much  closer to $\sqrt{N_{\rm O}}$ because the SNRs of individually detectable BNSs are more concentrated around $\rho_{\rm thr}$
and therefore the bias term of each merger $|\sigma_h(f;{\bf\Theta}^{\rm ML}(d))|^2_i\propto \rho_i^{-2}$
is of similar magnitude.
As a result, the fractional residual energy of BNSs turns out to be $\delta_2^{\rm rfn}\Omega/\Omega_{\rm GW}\sim \sqrt{N_{\rm O}}/\sum_i\rho_i^{2} $, which is lower than the residual of Method 1 $\delta_1^{\rm rfn}\Omega/\Omega_{\rm GW}\sim \sum_i\rho_i^{-1}/\sum_i\rho_i^{2}$
(see Fig.~\ref{fig:app2}).

\subsection{Sensitivity to the Background}
With either foreground cleaning method, we find the foreground from loud merger signals with SNR above the
detection threshold can be cleaned with residue energy density below the detector sensitivity limit $\Omega_{\rm det.lim.}$. As a result, the residual foreground is expected to be dominated by sub-threshold BNSs \cite{Regimbau2017,Sachdev2020,Zhong2022}),
which will be the next critical problem to solve for detecting the primordial SGWB in the 3G era.
This problem can be alleviated if
the design sensitivity of 3G detectors can be further improved, i.e., more BNSs would become individually
detectable if the detectors were more sensitive and the resulting unresolved foreground energy density $\Omega_{\rm unr}$
would be suppressed. A rough estimate shows that the unresolved BNS foreground
energy density $\Omega_{\rm unr}$ can be reduced by $\mathcal O(10^2)$ if all the 3G detector noise levels $\sqrt{P_{\rm n}(f)}$ were 3 times lower than what has been assumed in this work (see Fig.~\ref{fig:N_snr}). On the other hand,
it is possible to measure the unresolved BNS foreground $\Omega_{\rm unr}$
via the BNS merger rate at high redshift if the delay time between BNS mergers and BBH mergers is known.

If the BNS foreground can be cleaned with residue below the detector sensitive limit as in the BBH case,
the detector sensitivity to the primordial SGWB in the 3G era will be defined by $\Omega_{\rm det.lim.}(f)$ in
Fig.~\ref{fig:app1} or  \ref{fig:app2}. Taking a flat-spectrum SGWB as an example, a SGWB with energy density
$\Omega_{\rm SGWB}(f) =  {\rm min.} \{ \Omega_{\rm det.lim.}(f)\} = 10^{-13}$ is  detectable
 at 3 $\sigma$ confidence level with 4 yr observations in this optimistic case.
But if the unresolved BNS foreground $\Omega_{\rm unr}$ cannot be cleaned eventually,
the SGWB search and the residue foreground estimation must be simultaneous done \cite{Martinovic2021},
the detector sensitivity to the primordial SGWB in the 3G era will be defined
by $\Omega_{\rm unr}(f)+\Omega_{\rm det.lim.}(f)$ (see Fig.~\ref{fig:app2}).
A much louder SGWB with  energy density
$\Omega_{\rm SGWB}(f) = {\rm min.} \{\Omega_{\rm unr}(f)+ \Omega_{\rm det.lim.}(f)\}\approx 5\times 10^{-12}$ is  detectable
 at 3 $\sigma$ confidence level with 4 yr observations  in this pessimistic case.

\section{Discussion and Summary}

\subsection{Foreground Cleaning Complications: Parameter Inference Biases}

In the main text, we have not dealt with the impact of the SGWB on the binary parameter inference.
Taking the reference flat-spectrum SGWB as an example, it has little impact on the merger SNR with $P_{\rm n, SGWB}(f)/ P_{\rm n}(f) < 10^{-4}$
across the whole frequency range. But the binary parameter inference might be biased by the SGWB depending on how the detector noise
PSDs are measured, and we are to investigate its impact on the foreground cleaning in this subsection.

With the traditional method of measuring detector noise PSD from off-source data segments, the measured
detector noise is in fact the summation of the noise and the background, since the the SGWB is always on,
i.e., the true measured quantity is an effective noise PSD, $P_{\rm n}^{\rm eff}(f) = P_{\rm n}(f)+P_{\rm n, SGWB}(f)$,
with the two components unknown individually. As a result, the (effective) noises are correlated across different detectors due to the common SGWB component,
and the parameter inference is biased if the correlation is not correctly taken into account. In this case, the joint likelihood (for a 2-detector case) is formulated as
\be
\begin{aligned}
&-2\log L(d|{\bf\Theta})  \\
&=\int df\ (d_{(k)}- h_{(k)})\begin{bmatrix}
P_{\rm n,(1)} & \mathcal{R} P_{\rm n, SGWB} \\
\mathcal{R} P_{\rm n, SGWB}  & P_{\rm n,(2)}
\end{bmatrix}^{-1} (d_{(k)}^*- h_{(k)}^*)^{\rm T}\ , \\
&\approx \sum_k \braket{d_{(k)}- h_{(k)}|d_{(k)}- h_{(k)}}\\
&-2 \int df \frac{\mathcal{R} P_{\rm n, SGWB}}{P_{\rm n,(1)}P_{\rm n,(2)}} {\rm Real}\left[ (d_{(1)}- h_{(1)})(d_{(2)}- h_{(2)})^* \right]\ ,
\end{aligned}
\ee
where $\mathcal {R} \leq 1$ is the overlap reduction function between two detectors, and we have used the fact $P_{\rm n, SGWB}\ll P_{\rm n, (k)}$
in the approximate equal sign. The final line is the contribution to the likelihood from the SGWB, which is correlated among different detectors.
Using the same technique in subsection~\ref{subsec:previous}, we find the SGWB induces a parameter inference shift
$\delta\Theta_{\rm SGWB}\sim P_{\rm n, SGWB}/\sqrt{ P_{\rm n, (1)} P_{\rm n, (2)}} \delta\Theta_{\rm noise} < 10^{-4}  \delta\Theta_{\rm noise} $,
and   $\delta\Theta_{\rm SGWB}$ is correlated with $ \delta\Theta_{\rm noise}$.
As a result, the SGWB introduce a bias term to the residue PSD $|h_{,\alpha}\delta\Theta^{\alpha}_{\rm noise}h_{,\beta}\delta\Theta^{\beta}_{\rm SGWB}|
< 10^{-4}|h_{,\alpha}\delta\Theta^{\alpha}_{\rm noise}|^2 \approx 10^{-4}\delta_1 H(f)$, which is orders of magnitude lower than the detector
sensitivity limit therefore is safe to ignore.

Considering the abundance of merger signals, especially BNSs, the traditional method of measuring detector noise PSD from off-source data segments might be challenging.
For ET-like triangle-shape detectors, the signal-free stream by summing the strain outputs from the three interferometers has been shown to be useful
in measuring the detector noise PSD \cite{Wong2022,Goncharov2022,Janssens2022}.
With this method, the detector noise PSD $P_{\rm n}(f)$ can be measured separately from the SGWB. Using a similar likelihood analysis as in Subsection~\ref{subsec:previous},
we find  ${\bf\Theta}^{\rm ML} ={\bf\Theta}^{\rm true}+\delta{\bf\Theta}_{\rm noise}+\delta{\bf\Theta}_{\rm SGWB}$,
with $\delta\Theta_{\rm noise}^\alpha = (F^{-1})^{\alpha\beta} \braket{h_{,\beta}|n}$,
and $\delta\Theta_{\rm SGWB}^\alpha = (F^{-1})^{\alpha\beta} \braket{h_{,\beta}|n_{\rm SGWB}}$ \cite{Antonelli2021}. In this case,
$\delta\Theta^{\rm noise}$ and $\delta\Theta^{\rm SGWB}$ are uncorrelated.
As a result, the SGWB introduce a bias term to the residue PSD $|h_{,\alpha}\delta\Theta^\alpha_{\rm SGWB}|^2
\sim |h_{,\alpha}\delta\Theta_\alpha^{\rm noise}|^2 \times P_{\rm n, SGWB}/P_{\rm n}
\sim \delta_1 H(f) \times P_{\rm n, SGWB}/P_{\rm n} < 10^{-4}  \delta_1 H(f)$, which is safe to ignore too.

Another potential source of parameter inference bias is the calibration uncertainty of detector noise PSD. Assuming a constant calibration uncertainty $s^{\rm cal}$ (no frequency
dependence), i.e., $P_{\rm n} = (1+s^{\rm cal})P_{\rm n}^{\rm true}$, the binary parameter is biased by
$\delta\Theta_{\rm cal} \approx s^{\rm cal} \delta\Theta_{\rm noise} $, the resulting bias in the residue PSD
$\delta^{\rm cal}H(f)= |h_{,\alpha}\delta\Theta_{\rm noise}^\alpha h_{,\beta}\delta\Theta_{\rm cal}^\beta|\approx s^{\rm cal}\delta_1 H(f)$.
As a result,
the calibration uncertainty $s^{\rm cal}$ must be less than $10\%$ requiring that $\delta^{\rm cal}\Omega (f) < \delta \Omega_{\rm det.lim}$ for both BBHs and BNSs (see Figs.~\ref{fig:app1},\ref{fig:app2}).

\subsection{Summary}
To better probe  the primordial SGWB from early universe processes, the astrophysical foreground from
compact binary mergers must be cleaned, ideally with the residue foreground energy density below the detector sensitive limit $\Omega_{\rm res}(f) < \Omega_{\rm det.lim.}(f)$. A number of subtraction methods have been proposed for cleaning the
foreground from individually detectable merger signal with $\rho > \rho_{\rm thr}$.
In terms of fractional residue after foreground cleaning, the state of the art subtraction method proposed in \cite{Sachdev2020}
is of $\delta H/H\sim \mathcal O(1)$ \cite{Zhou2022,Zhou2022b}, and the projection method proposed in \cite{Cutler2006,Harms2008} is of $\delta H/H\sim \rho^{-2}$ \cite{Sharma2020}.
 For the cleaning method in the time-frequency domain proposed in \cite{Zhong2022}, it is not straightforward to quantify the fractional residue scaling with $\rho$.
With these methods, it is
not sufficient to reduce foreground to the target level $\Omega_{\rm det.lim.}(f)$. In this work, we  proposed a foreground cleaning method  by first subtracting  the signal strain from data using
the ML strain as a proxy, then removing the average residual power,  and it turns out that Method 1 is of $\delta H/H\sim \rho^{-3}$ (and the alternative method by measuring the foreground PSD only is of $\delta H/H\sim \rho^{-2}/4$ for BBHs and $\sim \rho^{-2}/\sqrt{N_{\rm O}}$ for BNSs).  Simulations under simplified assumptions show that these two methods
are sufficient to reduce the foreground from individually detectable binary mergers to the target level.

However, the unresolved foreground from subthrehold BNSs is expected to dominate the residual foreground
\cite[see also][]{Regimbau2017,Sachdev2020,Zhong2022}, which will be the next
critical problem to solve for detecting the primordial SGWB in the 3G era.
If the unresolved BNS foreground can be effectively cleaned to below the detector sensitivity level,
a flat SGWB with $\Omega_{\rm SGWB}=10^{-13}$ is expected to be detected at 3 $\sigma$ confidence level with 4 yr observations.
If the unresolved BNS foreground cannot
be efficiently cleaned or accurately measured, it will be a bottle neck for detecting the primordial SGWB,
and only a much louder flat SGWB with $\Omega_{\rm SGWB}=5\times 10^{-12}$  can
be detectable at the same confidence level with the same observation time.

Current foreground methods have only been tested in idealized simulations,
where there is no overlapping merger signals and detector noise is Gaussian with known PSD.
In fact, the overlapping merger signals \cite{Himemoto2021,Pizzati2022} make the signal parameter inference more complicate and possibly biased if not well taken care of,
and the traditional method of measuring detector noise PSD from off-source data segments might be challenging
due to the abundance of merger signals.
For ET-like triangle-shape detectors, the signal-free stream by summing the strain outputs from the three interferometers has been shown to be useful
in measuring the detector noise PSD \cite{Wong2022,Goncharov2022,Janssens2022}.
For L-shape detectors, the measurement uncertainty of detector noise PSD is expected to be higher.
The detector noise calibration uncertainty must be less than $10\%$ requiring that the resulting residue PSD bias is lower than the detected sensitivity limit.

\vspace{0.2 cm}

\emph{Acknowledgement.} We thank Liang Dai, Neal Dalal, Reed Essick and Junwu Huang for valueable discussions.
We also thank Bei Zhou for sharing the detector sensitivity limit curve.
We are supported by the Natural Sciences
and Engineering Research Council of Canada and in part
by Perimeter Institute for Theoretical Physics. Research at
Perimeter Institute is supported in part by the Government of
Canada through the Department of Innovation, Science and
Economic Development Canada and by the Province of Ontario through the Ministry of Colleges and Universities.

\bibliography{ms}

\end{document}